\documentclass[english,aps,prb,showpacs,superscriptaddress,floats,amsmath,amssymb,floatfix,dblfloatfix,nobalancelastpage,twocolumn]{revtex4-1}
\usepackage[T1]{fontenc}
\usepackage[latin9]{inputenc}
\usepackage{amsmath}
\usepackage{amssymb}
\usepackage{graphicx}
\usepackage{babel}
\begin{document}

\title{Phonon renormalization in the Kitaev quantum spin liquid}

\author{Alexandros Metavitsiadis}

\affiliation{Institute for Theoretical Physics, Technical University Braunschweig,
D-38106 Braunschweig, Germany}

\author{Wolfram Brenig}

\affiliation{Institute for Theoretical Physics, Technical University Braunschweig,
D-38106 Braunschweig, Germany}
\begin{abstract}
We study the self-energy of phonons, magnetoelastically coupled to
the two-dimensional Kitaev spin-model on the honeycomb lattice. Fractionalization
of magnetic moments into mobile Majorana matter and a static $\mathbb{Z}_{2}$
gauge field lead to a continuum of relaxation processes comprising
two channels. Thermal flux excitations, which act as an emergent disorder,
strongly affect the phonon renormalization. Above the flux proliferation
temperature, the dispersion of a narrow quasiparticle-hole channel
is suppressed in favor of broad and only weakly momentum dependent
features, covering large spectral ranges. Our analysis is based on
complementary calculations in the low-temperature homogeneous gauge
and a mean-field treatment of thermal gauge fluctuations, valid at
intermediate and high temperatures.
\end{abstract}
\maketitle

\section{Introduction}

Quantum spin liquids (QSL) are intriguing forms of matter, in which
local magnetic order parameters are absent even at zero temperature.
QSLs can result from frustrated magnetic exchange and may show many
peculiar properties, which are of great current interest. Among them
are fractionalized excitations, topological entanglement, and quantum
orders\cite{Balents2010,Balents2016}. Many models have been proposed,
to approximately exhibit QSL behavior. Kitaev's compass exchange Hamiltonian
on the honeycomb lattice is one of the few, in which a $\mathbb{Z}_{2}$
QSL can exactly be shown to exist\cite{Kitaev2006}. The spin degrees
of freedom of this model fractionalize in terms of mobile Majorana
fermions coupled to a static $\mathbb{Z}_{2}$ gauge field\cite{Kitaev2006,Feng2007,Chen2008,Nussinov2009,Mandal2012}.
Mott-insulators with strong spin-orbit coupling (SOC) may be a fertile
ground for Kitaev materials\cite{Khaliullin2005,Jackeli2009,Chaloupka2010,Nussinov2015}.
However, residual non-Kitaev exchange interactions remain an issue,
driving most of the present systems into magnetic order at low temperatures\cite{Trebst2017}.

Free mobile Majorana fermions have been invoked to interpret ubiquitous
unconventional continua in spectroscopies on Kitaev materials, like
inelastic neutron\cite{Banerjee2016,Banerjee2016a,Banerjee2018} and
Raman scattering\cite{Knolle2014}, as well as local resonance probes\cite{Baek2017,Zheng2017}.
Majorana fermions may also play a role in thermal transport. Here,
$\alpha$-RuCl$_{3}$ \cite{Plumb2014} has been under intense scrutiny.
A \emph{transverse} thermal conductivity $\kappa_{xy}$ in magnetic
fields, i.e. a thermal Hall effect, and its potential quantization
has been observed\cite{Kasahara2018}. This may be an evidence for
chiral Majorana edge modes. Alternative explanations in terms of chiral
magnon edge states have been given\cite{Cookmeyer2018,McClarty2018},
lacking quantization of $\kappa_{xy}$ however.

In any real Kitaev system, proximate to a QSL, the Majorana fermions
will be subject to unavoidable perturbations, including e.g. non-Kitaev
exchange, defects and coupling to lattice degrees of freedom. While
the former two have received considerable attention, spin-phonon coupling
in Kitaev magnets remains to be explored. Phonon-Majorana mixing has
been shown to degrade the thermal Hall plateaus\cite{Ye2018,Vinkler-Aviv2018}.
For the \emph{longitudinal} thermal conductivity $\kappa_{xx}$ in
$\alpha$-RuCl$_{3}$\cite{Hirobe2017,Leahy2017,Hentrich2018,Yu2018},
a picture has emerged where heat transport is primarily governed by
phonons and phonon-Majorana scattering has been suggested to be an
important dissipation mechanism\cite{Hentrich2018,Yu2018}. Various
other indications of phonons mixing with putative Majorana particles
in $\alpha$-RuCl$_{3}$ have been reported in Raman scattering\cite{Glamazda2017,Sahasrabudhe2019}
optical absorption\cite{Reschke2019}, and thermodynamic measurements\cite{Widmann2019}.
Finally, magnetoelastic coupling along the Ru-Ru links in $\alpha$-RuCl$_{3}$
is known to be significant, driving a transition into a pressure induced
valence bond state\cite{Biesner2018,Bastien2018,Yadav2018}.

In this context, the main purpose of this work is to uncover signatures
of Majorana fermions in phonon spectra of Kitaev magnets. We focus
on the the long wave-length limit of acoustic modes, which play an
important role in thermal transport. We find that the combination
of a fermionic Dirac-cone spectrum and the thermal excitations of $\mathbb{Z}_{2}$
gauge fluxes lead to dramatic deviations of the phonon self-energies
as compared to conventional phonon-electron/magnon scattering. The
outline of the paper is as follows. In Sec. \ref{sec:mpc} we describe
the microscopic model. Sec. \ref{sec:sigma} details our calculations,
discriminating between the low-temperature gauge ground state in Subsec.
\ref{subsec:homg} and the flux-proliferated state at elevated temperatures
in Subsec. \ref{subsec:rand}. In Sec. \ref{sec:Results} results
for the phonon self-energy are discussed. Finally, we summarize in
Sec. \ref{sec:Summary}.

\section{Spin-phonon coupling\label{sec:mpc}}

We consider the Kitaev spin-model on the two dimensional honeycomb
lattice
\begin{equation}
H_{0}=\sum_{{\bf l},\alpha}J_{\alpha}S_{{\bf l}}^{\alpha}S_{{\bf l}+{\bf r}_{\alpha}}^{\alpha}\,,\label{eq:1}
\end{equation}
where ${\bf l}=n_{1}{\bf R}_{1}+n_{2}{\bf R}_{2}$ runs over the sites
of the triangular lattice with ${\bf R}_{1[2]}=(1,0),\,[(\frac{1}{2},\frac{\sqrt{3}}{2})]$,
and ${\bf r}_{\alpha=x,y,z}=(\frac{1}{2},\frac{1}{2\sqrt{3}}),$ $(-\frac{1}{2},\frac{1}{2\sqrt{3}})$,
$(0,-\frac{1}{\sqrt{3}})$ refer to the basis sites $\alpha=x,y,z$,
tricoordinated to each lattice site of the honeycomb lattice. As is
well documented in the literature \cite{Kitaev2006}, (\ref{eq:1})
can be mapped onto a quadratic form of Majorana fermions in the presence
of a static $\mathbb{Z}_{2}$ gauge $\eta_{{\bf l}}=\pm1$, residing
on, e.g., the $\alpha=z$ bonds
\begin{equation}
H_{0}=-\frac{i}{2}\sum_{{\bf l},\alpha}J_{\alpha}\eta_{{\bf l},\alpha}\,a_{{\bf l}}c_{{\bf l}+{\bf r}_{\alpha}}\,,\label{eq:2}
\end{equation}
where $\eta_{{\bf l},\alpha}$ is introduced to unify the notation,
with $\eta_{{\bf l},x(y)}=1$ and $\eta_{{\bf l},z}=\eta_{{\bf l}}$.
There are two types of Majorana particles, corresponding to the two
basis sites. We chose to normalize them as $\{a_{{\bf l}},a_{{\bf l}'}\}=\delta_{{\bf l},{\bf l}'}$,
$\{c_{{\bf m}},c_{{\bf m}'}\}=\delta_{{\bf m},{\bf m}'}$, and $\{a_{{\bf l}},c_{{\bf m}}\}=0$.
For each gauge sector $\{\eta_{{\bf l}}\}$, (\ref{eq:2}) represents
a spin liquid.

Various types of phonon couplings to (pseudo-)spins in SOC matter
can be invoked microscopically, including one- and two-spin processes.
Here we focus on lattice deformations ${\bf u}_{{\bf l}}$ introduced
into (\ref{eq:1}) by a magnetoelastic coupling approach, i.e. $J_{\alpha}$
changes to $J_{\alpha}+\nabla J_{\alpha}\cdot({\bf u}_{{\bf l}+{\bf r}_{\alpha}}-{\bf u}_{{\bf l}})$.
This kind of bond dependent modification of the exchange leave the
mapping from (\ref{eq:1}) to (\ref{eq:2}) intact, i.e. the magnetoelastic
coupling can be considered directly on the level of the Majorana hamiltonian
\begin{equation}
H=H_{0}-\frac{i}{2}\sum_{{\bf l},\alpha}\nabla J_{\alpha}\cdot({\bf u}_{{\bf l}+{\bf r}_{\alpha}}-{\bf u}_{{\bf l}})\,\eta_{{\bf l},\alpha}\,a_{{\bf l}}c_{{\bf l}+{\bf r}_{\alpha}}\,,\label{eq:3}
\end{equation}
To simplify, we set $\nabla J_{\alpha}\approx\lambda{\bf r}_{\alpha}$,
i.e. bond-'strechting' is assumed to be the primary source of spin-lattice
coupling. Quantizing the deformations into phonons, we confine the
analysis to the long wave-length limit of the acoustic spectrum. Due
to this, we may discard the non-Bravais nature of the honeycomb lattice
for the phonons\cite{Stamokostas2017} and introduce only a \textit{single}
type of boson $b_{{\bf q}\mu}^{(\dagger)}$\cite{Stamokostas2017}
\[
{\bf u}_{{\bf l}}=\frac{1}{\sqrt{N}}\sum_{{\bf q}}\frac{{\bf P}_{{\bf q}\mu}}{\sqrt{2m\omega_{{\bf q}\mu}}}(b_{{\bf q}\mu}^{\phantom{\dagger}}+b_{-{\bf q}\mu}^{\dagger})\,e^{i{\bf q}\cdot{\bf l}}
\]
with momentum ${\bf q}$, \emph{normalized} polarization vectors ${\bf P}_{{\bf q}\mu}$,
with index $\mu$ for longitudinal or transverse modes, effective ionic
mass $m$, and phonon dispersion $\omega_{{\bf q}\mu}$. 2D phonon
momenta are assumed for the remainder of this work. With this, the
Majorana-phonon coupling, i.e. the 2nd term in (\ref{eq:3}) reads
\begin{align}
H_{MP} & =\frac{1}{\sqrt{N}}\sum_{{\bf q}\mu}(b_{{\bf q}\mu}^{\phantom{\dagger}}{+}b_{-{\bf q}\mu}^{\dagger})V_{{\bf q}\mu}\label{eq:4}\\
V_{{\bf q}\mu} & =\sum_{{\bf l},\alpha}\frac{-i\lambda{\bf P}_{{\bf q}\mu}\cdot{\bf r}_{\alpha}}{2\sqrt{2m\omega_{{\bf q}\mu}}}(e^{i{\bf q}\cdot\boldsymbol{r}_{\alpha}}-1)\eta_{{\bf l},\alpha}e^{i{\bf q}\cdot{\bf l}}a_{{\bf l}}c_{{\bf l}+{\bf r}_{\alpha}}\,,\nonumber 
\end{align}
which is Hermitian, i.e. $V_{-{\bf q}\mu}^{\phantom{\dagger}}=V_{{\bf q}\mu}^{\dagger}$.

\section{Phonon self-energy\label{sec:sigma}}

In this section we present our evaluation of the phonon self-energy.
We focus on two temperature regimes, namely $T\lesssim(\gtrsim)T^{\star}$.
Here $T^{\star}$ is the so called flux proliferation temperature.
In the vicinity of this temperature the gauge field and therefore
fluxes get thermally excited. Previous analysis\cite{Nasu2015,Metavitsiadis2017,Pidatella2019}
has shown, that the temperature range over which a \emph{complete}
proliferation of fluxes occurs is confined to a rather narrow region,
less than a decade centered around $T^{\star}\approx0.012J$ for isotropic
exchange, $J$=$J_{x,y,z}$, used in this work, and decrease rapidly with anisotropy\cite{Nasu2015,Pidatella2019}.
Our strategy therefore is to consider a homogeneous ground state gauge,
i.e. $\eta_{{\bf l}}=1$ for $T\lesssim T^{\star}$ and completely
random-gauge states for $T\gtrsim T^{\star}$. This approach has proven
to work very well on a \emph{quantitative} level in several studies
of the thermal conductivity of Kitaev models\cite{Metavitsiadis2017,Pidatella2019,Metavitsiadis2017a}.

\subsection{Homogeneous gauge for $T\lesssim T^{\star}$\label{subsec:homg}}

For $\eta_{{\bf l}}=1$ the Hamiltonian (\ref{eq:1}) can be diagonalized
\emph{analytically} in terms of complex Dirac fermions. Mapping from
the real Majorana fermions to the latter can be achieved in various
ways, all of which require some type of linear combination of real
fermions in order to form complex ones. Here we do the latter by using
Fourier transformed Majorana particles, $a_{{\bf k}}^{\phantom{\dagger}}=\sum_{{\bf l}}e^{-i{\bf k}\cdot{\bf l}}a_{{\bf l}}/\sqrt{N}$
with momentum ${\bf k}$ and analogously for $c_{{\bf k}}^{\phantom{\dagger}}$.
The prime reason for this is to remain with the structure of the honeycomb
lattice. Other popular approaches\cite{Feng2007,Nussinov2015}, lead
to effective lattices which may pose issues regarding the discrete
rotational symmetry of the phonon self-energy. 

The fermions introduced in momentum space are complex with $a_{{\bf k}}^{\dagger}=a_{-{\bf k}}^{\phantom{\dagger}}$,
i.e. only half of the momentum states are independent. This rephrases,
that for each Dirac fermion, there are two Majorana particles. Standard
anticommutation relations apply, $\{a_{{\bf k}}^{\phantom{\dagger}},a_{{\bf k}'}^{\dagger}\}=\delta_{{\bf k},{\bf k}'}$,
$\{c_{{\bf k}}^{\phantom{\dagger}},c_{{\bf k}'}^{\dagger}\}=\delta_{{\bf k},{\bf k}'}$,
and $\{a_{{\bf k}}^{(\dagger)}c_{{\bf k}'}^{(\dagger)}\}=0$. Using
this, the diagonal form of $H$ reads 
\begin{equation}
H=\sum_{{\bf k},\gamma=1,2}^{\sim}\mathrm{sg}_{\gamma}\,\epsilon_{{\bf k}}\,d_{{\bf k},\gamma}^{\dagger}d_{{\bf k},\gamma}^{\phantom{\dagger}}\,,\label{eq:5}
\end{equation}
where the $\tilde{\sum}$ sums over a reduced 'positive' half of momentum
space and $\mathrm{sg}_{\gamma}$=1(-1) for $\gamma$=1(2). The quasiparticle
energy is $\epsilon_{{\bf k}}=J[3+2\cos(k_{x})+4\cos(k_{x}/2)\cos(\sqrt{3}k_{y}/2)]^{1/2}/2$.
In terms of reciprocal lattice coordinates $x,y\in[0,2\pi]$, this
reads $\epsilon_{{\bf k}}=J[3+2\cos(x)+2\cos(y)+2\cos(x-y)]^{1/2}/2$
with ${\bf k}=x\,{\bf G}_{1}+y\,{\bf G}_{2}$, where ${\bf G}_{1[2]}=(1,-\frac{1}{\sqrt{3}}),\,[(0,\frac{2}{\sqrt{3}})]$.
The quasiparticles are given by
\begin{align}
\left[\begin{array}{c}
c_{{\bf k}}\\
a_{{\bf k}}
\end{array}\right] & =\left[\begin{array}{cc}
u_{11}({\bf k}) & u_{12}({\bf k})\\
u_{21}({\bf k}) & u_{22}({\bf k})
\end{array}\right]\left[\begin{array}{c}
d_{1{\bf k}}\\
d_{2{\bf k}}
\end{array}\right]\label{eq:6}\\
u_{11}({\bf k}) & =-u_{12}({\bf k})=\frac{i\sum_{\alpha}e^{-i{\bf k}\cdot{\bf r}_{\alpha}}}{2^{3/2}\epsilon_{{\bf k}}}\nonumber \\
u_{21}({\bf k}) & =u_{22}({\bf k})=\frac{1}{\sqrt{2}}\,.\nonumber 
\end{align}
From the sign change of the quasiparticle energy between bands $\gamma$=1,2
in Eq. (\ref{eq:5}) it is clear that the relations $a_{{\bf k}}^{\dagger}=a_{-{\bf k}}^{\phantom{\dagger}}$ and 
$c_{{\bf k}}^{\dagger}=c_{-{\bf k}}^{\phantom{\dagger}}$
for reversing momenta of the original Majorana ferminons has to change
into $d_{1(2){\bf k}}^{\dagger}=d_{2(1)-{\bf k}}^{\phantom{\dagger}}$,
switching also the bands. Indeed this is also born out of the transformation
(\ref{eq:6}). The phonon quasiparticle vertex $V_{{\bf q}\mu}$ from
Eq. (\ref{eq:4}) turns into
\begin{align}
V_{{\bf q}\mu} & =\frac{1}{N}\sum_{{\bf k}}g_{{\bf k},{\bf q},\mu}\left[u_{11}^{\star}({\bf k}{+}{\bf q})(d_{1{\bf k}+{\bf q}}^{\dagger}d_{1{\bf k}}^{\phantom{\dagger}}+d_{1{\bf k}+{\bf q}}^{\dagger}d_{2{\bf k}}^{\phantom{\dagger}})\right.\nonumber \\
 & \left.\phantom{aaaaaaa}+u_{12}^{\star}({\bf k}{+}{\bf q})(d_{2{\bf k}+{\bf q}}^{\dagger}d_{1{\bf k}}^{\phantom{\dagger}}+d_{2{\bf k}+{\bf q}}^{\dagger}d_{2{\bf k}}^{\phantom{\dagger}})\right]\nonumber \\
g_{{\bf k},{\bf q},\mu} & =\frac{i\lambda{\bf P}_{{\bf q}\mu}\cdot{\bf r}_{\alpha}}{2^{3/2}\sqrt{2m\omega_{{\bf q}\mu}}}(1-e^{-i{\bf q}\cdot\boldsymbol{r}_{\alpha}})e^{-i{\bf k}\cdot{\bf r}_{\alpha}}\,.\label{eq:7}
\end{align}
To obtain the phonon renormalization we evaluate the self-energy $\Sigma_{\mu\nu}(q,i\omega_{n})$
of the boson propagator $\langle T_{\tau}(b_{{\bf q}\mu}^{\phantom{\dagger}}(\tau)b_{{\bf q}\nu}^{\dagger})\rangle\stackrel{\tau\rightarrow i\omega_{n}}{=}[(i\omega_{n}-\omega_{{\bf q}\mu})\delta_{\mu\nu}-\Sigma_{\mu\nu}({\bf q},i\omega_{n})]^{-1}$. Renormalized phonon energies $z_{\bf q}$ follow from the secular equation $\mathrm{det}\{[z_{\bf q}^{2}-\omega_{{\bf q}\mu}^{2}]\delta_{\mu\nu}-2\omega_{{\bf q}\mu}\Sigma_{\mu\nu}({\bf q},z_{\bf q})\}=0$\cite{AGD}. 
To make progress, we proceed by perturbation theory to $O(V^{2}$).
This leaves aside potential concerns about Migdal's theorem\cite{Migdal1958,Roy2014}.
In order to ease geometrical complexity, we refrain from confining
the complex fermions to only a reduced ``positive'' region of momentum
space. This comes at the expense of additional anomalous anticommutators
like e.g. $\{d_{1{\bf k}}^{\phantom{\dagger}},d_{2{\bf k}'}^{\phantom{\dagger}}\}=\delta_{-{\bf k},{\bf k}'}$
and their corresponding contractions. After some algebra, we find
\begin{align}
\Sigma_{\mu\nu}({\bf q},z) & =\Sigma_{\mu\nu}^{ph}({\bf q},z)+\Sigma_{\mu\nu}^{pp}({\bf q},z)\label{eq:8}\\
\Sigma_{\mu\nu}^{ph}({\bf q},z) & =\frac{1}{N}\sum_{{\bf k}}A_{k,q,\mu}^{ph}A_{k,q,\nu}^{ph\star}\frac{f_{{\bf k}+{\bf q}}(T){-}f_{{\bf k}}(T)}{z-\epsilon_{{\bf k}+{\bf q}}+\epsilon_{{\bf k}}}\nonumber \\
\Sigma_{\mu\nu}^{pp}({\bf q},z) & =\frac{1}{2N}\sum_{{\bf k}}A_{k,q,\mu}^{pp}A_{k,q,\nu}^{pp\star}\left\{ [f_{{\bf k}+{\bf q}}(T){+}f_{{\bf k}}(T){-}1]\right.\nonumber \\
 & \phantom{aaa}\times\left.\left(\frac{1}{z-\epsilon_{{\bf k}+{\bf q}}-\epsilon_{{\bf k}}}{+}\frac{1}{-z-\epsilon_{{\bf k}+{\bf q}}-\epsilon_{{\bf k}}}\right)\right\} \,,\nonumber 
\end{align}
where the superscripts $ph(pp)$ indicate particle-hole(particle-particle)
type of intermediate states of the Dirac fermions, $f_{{\bf k}}(T)=1/(e^{\epsilon_{{\bf k}}/T}+1)$
is the Fermi function, $z\in\mathbb{C}$ with $\mathrm{Im}(z)>0$,
and the transition matrix elements are
\begin{equation}
A_{k,q,\mu}^{\stackrel{{\scriptstyle ph}}{pp}}=g_{{\bf k},{\bf q},\mu}u_{11}^{\star}({\bf k}{+}{\bf q})\pm g_{-{\bf k}-{\bf q},{\bf q},\mu}u_{11}^{\star}(-{\bf k})\,,\label{eq:9}
\end{equation}
where the ${+}$(${-}$) sign corresponds to the $ph$($pp$) channel.
This concludes the formal details for $T\lesssim T^{\star}$.

\subsection{Random gauge for $T\gtrsim T^{\star}$\label{subsec:rand}}

In a random gauge configuration, translational invariance of the Majorana
system is lost, and we resort to a numerical approach in real space.
First a spinor $A_{\text{\ensuremath{\sigma}}}^{\dagger}=(a_{1}\dots a_{{\bf l}}\dots a_{N},c_{1}\dots c_{{\bf l}+{\bf r}_{x}}\dots c_{N})$,
comprising the Majoranas on the $2N$ sites of the lattice is defined.
Using this, Hamiltonian (\ref{eq:3}) is rewritten as $H={\bf A}^{\dagger}({\bf h}_{0}+{\bf h}_{MP}){\bf A}/2$.
Bold faced symbols refer to vectors and matrices, i.e. ${\bf h}_{0(MP)}$
are $2N\times2N$ arrays. Next a spinor $D_{\sigma}^{\dagger}=(d_{1}^{\dagger}\dots d_{N}^{\dagger},d_{1}^{\phantom{\dagger}}\dots d_{N}^{\phantom{\dagger}})$
of $2N$ complex fermions is defined by ${\bf D}={\bf F}{\bf A}$
using the unitary (Fourier) transform ${\bf F}$. The latter is built
from two disjoint $N\times N$ blocks $f_{\sigma\rho}^{i=1,2}=e^{-i{\bf k}_{\sigma}\cdot{\bf R}_{\rho}^{i}}/\sqrt{N}$,
with ${\bf R}_{\rho}^{i}={\bf l}$ and ${\bf l}+{\bf r}_{x}$, for
$a$- and $c$-Majorana lattice sites, respectively. ${\bf k}$ is
chosen such, that for each ${\bf k}$, there exists one $-{\bf k}$,
with ${\bf k}\neq-{\bf k}$. Finally, for convenience, ${\bf F}$
is rearranged such as to associate the $d_{1}^{\dagger}\dots d_{N}^{\dagger}$
with the $2\,(N/2)=N$ 'positive' ${\bf k}$-vectors. With this 
\begin{equation}
H={\bf D}^{\dagger}\,[{\bf \tilde{h}}_{0}+\frac{1}{\sqrt{N}}\sum_{{\bf q}\mu}(b_{{\bf q}\mu}^{\phantom{\dagger}}{+}b_{-{\bf q}\mu}^{\dagger}){\bf \tilde{v}}_{{\bf q}\mu}]\,{\bf D}/2\,,\label{eq:10}
\end{equation}
where $\tilde{{\bf o}}={\bf F}{\bf o}{\bf F}^{\dagger}$ and ${\bf v}_{{\bf q}\mu}$
stems from Eq. (\ref{eq:4}). We emphasize, that (i) ${\bf F}$ does
\emph{not} diagonalize $H$ and (ii) that in general, the $2N\times2N$
matrices of Fourier transformed operators $\tilde{{\bf o}}$ will
contain particle number non-conserving entries of ${\bf D}$ fermions.

As for the case of the homogeneous gauge in Sec. \ref{subsec:homg}
the phonon self-energy for a \emph{particular} gauge sector $\{\eta_{{\bf l}}\}$
is given by
\begin{equation}
\Sigma_{\mu\nu}({\bf q},\tau)=\frac{1}{4}\langle T_{\tau}[({\bf D}^{\dagger}{\bf \tilde{v}}_{{\bf q}\mu}{\bf D})(\tau)({\bf D}^{\dagger}{\bf \tilde{v}}_{{\bf q}\mu}{\bf D})^{\dagger}]\rangle_{\{\eta_{{\bf l}}\}}\,.\label{eq:11}
\end{equation}
This is evaluated using Wick's theorem for quasiparticles ${\bf T}={\bf U}{\bf D}$,
referring to a $2N\times2N$ Bogoliubov transformation ${\bf U}$
which is determined numerically for a given distribution $\{\eta_{{\bf l}}\}$
and which diagonalizes $({\bf U}\tilde{{\bf h}_{0}}{\bf U}^{\dagger})_{\rho\sigma}=\delta_{\rho\sigma}\epsilon_{\rho}$,
with $\epsilon_{\rho}=(\epsilon_{1}\dots\epsilon_{N},-\epsilon_{1}\dots-\epsilon_{N})$.
We get
\begin{align}
\Sigma_{\mu\nu}({\bf q},z) & =\sum_{\rho\sigma}\Pi_{\sigma\rho}(z)w_{\sigma\rho,{\bf q}\mu}[w_{\bar{\rho}\bar{\sigma},{\bf q}\nu}^{\star}-w_{\sigma\rho,{\bf q}\nu}^{\star}]\nonumber \\
\Pi_{\sigma\rho}(z) & =\frac{f_{\sigma}(T)-f_{\rho}(T)}{z-\epsilon_{\sigma}+\epsilon_{\rho}}\label{eq:12}\\
w_{\rho\sigma,{\bf q}\mu} & =(\frac{1}{2}{\bf U}\tilde{{\bf v}}_{{\bf q}\mu}{\bf U}^{\dagger})_{\rho\sigma}\,,\nonumber 
\end{align}
where $f_{\sigma}(T)=1/(e^{\epsilon_{\sigma}/T}+1)$, and overbars refer to swapping the upper and
lower half of the range of 2$N$ indices, e.g. $\bar\rho = \rho \mp N$ for $\rho \gtrless N $. For clarity
sake we note, that the indices $\mu,\nu$ refer to three phonon polarizations,
while the indices $\rho,\sigma$ label $2N$ quasiparticles.
\begin{figure}[tb]
\centering{}\includegraphics[width=0.95\columnwidth]{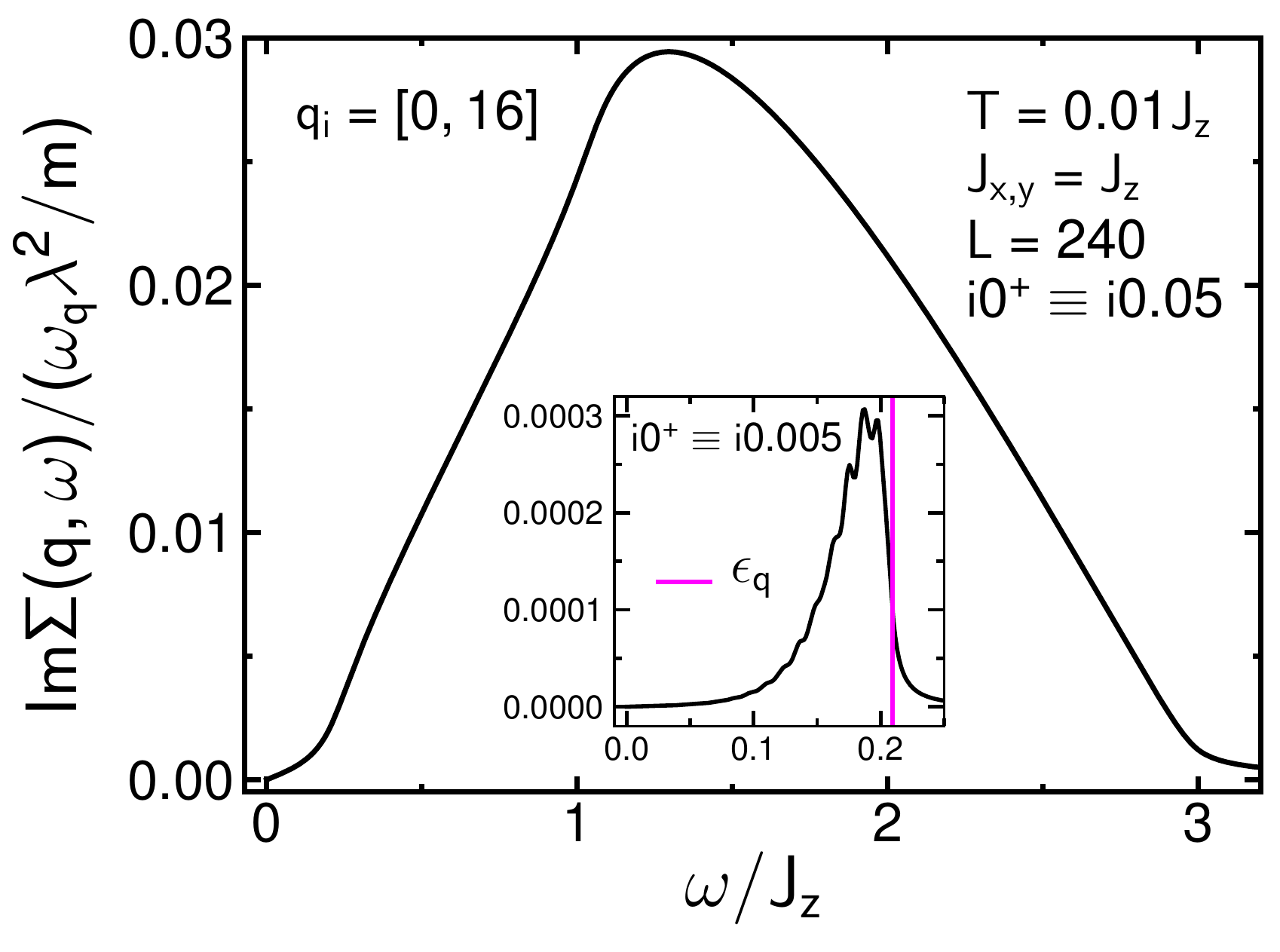}
\caption{Imaginary part, $\mathrm{Im}\Sigma({\bf q},\omega{+}i0^{+})$ of phonon
self-energy versus $\omega$ at fixed small ${\bf q}$, for low $T=0.01J_{z}\lesssim T^{\star}$,
using Eq. (\ref{eq:8}) for homogeneous ground state gauge. Momentum
${\bf q}=2\pi/L\sum_{j=1,2}q_{ij}{\bf G}_{j}$. Inset: Blow up of
low-$\omega$ region with reduced imaginary broadening. Magenta line:
upper ph-continuum bound at $T=0$.\label{fig:1}}
\end{figure}

As a final step, $\Sigma_{\mu\nu}({\bf q},z)$ from Eq. (\ref{eq:12})
is averaged over a sufficiently large number of random distributions
$\{\eta_{{\bf l}}\}$. This concludes the formal details of the evaluation
of the self-energy for $T\gtrsim T^{\star}$. 
\begin{figure*}[t]
\centering{}\includegraphics[width=1.82\columnwidth]{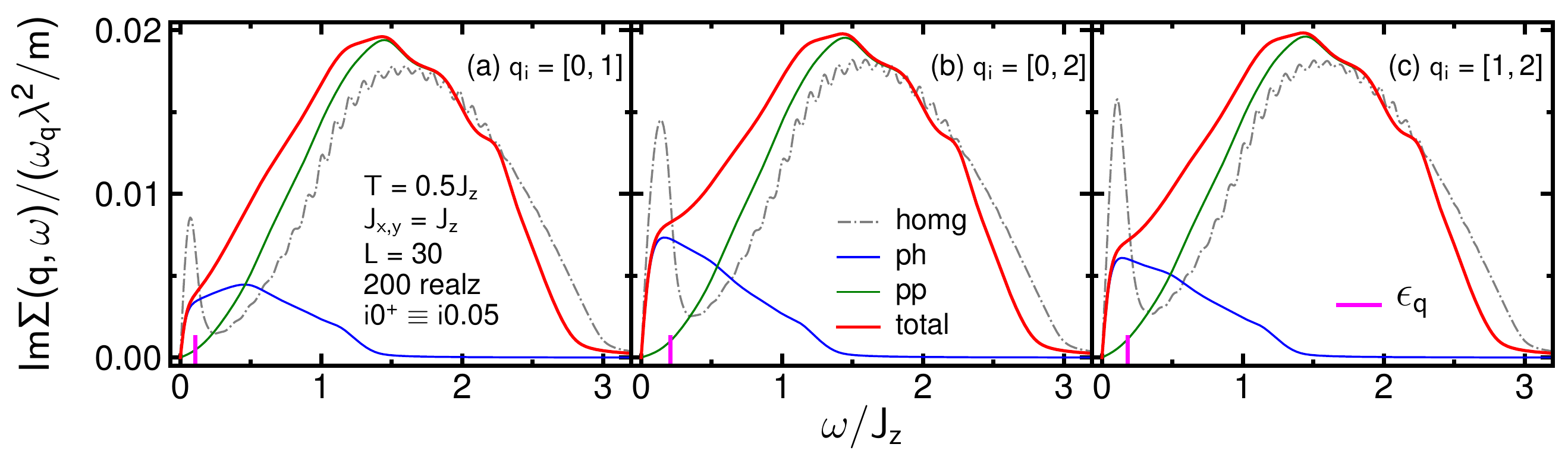}
\caption{Solid lines: Imaginary part, $\mathrm{Im}\Sigma({\bf q},\omega+i0^{+})$
of phonon self-energy at $T=0.5J_{z}\gtrsim T^{\star}$ versus $\omega$
for three fixed small momenta ${\bf q}=2\pi/L\sum_{j=1,2}q_{ij}{\bf G}_{j}$,
i.e. panels (a), (b), (c), using Eq. (\ref{eq:12}) for random gauge
state. Red, green, blue: total, ph, pp spectrum. Gray dashed dotted
line: $\mathrm{Im}\Sigma({\bf q},\omega+i0^{+})$ for identical system
parameters, however assuming homegeous ground state gauge. Magenta
line: upper ph-continuum bound at $T=0$.\label{fig:2}}
\end{figure*}

\section{Results\label{sec:Results}}

In this section we discuss selected properties of the phonon self-energy.
Ahead of that, several issues have to be addressed. First, we note
that material-specific analysis of phonons and a classification of
related (pseudo)-spin-phonon coupling processes in potential Kitaev
compounds has only begun recently. Most noteworthy, \emph{ab-initio}
calculations suggest that acoustic phonons in $\alpha$-RuCl$_{3}$
have sound velocities very similar to those of the fermions at the
Dirac cone\cite{Widmann2019,privateValenti2019}. In view of this,
we use a very simple phonon dispersion $\omega_{{\bf q}}=v_{p}[3-\cos(q_{x})-2\cos(q_{x}/2)\cos(\sqrt{3}q_{y}/2)]^{1/2}$,
meant solely to show some arbitrarily chosen form of sixfold symmetry
and we set $v_{p}\equiv1$, i.e. of $O(J_{x,y,z})$ hereafter. We
emphasize that $\omega_{{\bf q}}$ is simply a scale factor to $\Sigma_{\mu\nu}({\bf q},z)$
with only $\omega_{{\bf q}}\sim O({\bf q})$ at small ${\bf q}$ being
relevant for proper hydrodynamic behavior. Second, although $\Sigma_{\mu\nu}(q,i\omega_{n})$
is non-diagonal in principle, such mixing of different phonon branches, is not
expected to provide additional qualitative insight. Therefore, for the remainder
of this section we focus on a diagonal component $\Sigma({\bf q},z)\equiv\Sigma_{\mu\mu}({\bf q},z)$
of the self-energy, and we furthermore assume the corresponding polarization $\mu$
to be longitudinal. In any material-specific context, the meaning
of the latter may be intricate\cite{privateValenti2019}. Here we
set ${\bf P}$ to be a unit vector along ${\bf q}$. Third, all result
for $\Sigma({\bf q},z)$ are displayed in terms of $\Sigma({\bf q},z)/\omega_{{\bf q}}$,
which, in view of the Dyson equation for the phonons is the dimensionless
renormalization parameter of the phonon dispersion, i.e. $\omega_{{\bf q}}\rightarrow\omega_{{\bf q}}(1+\Sigma({\bf q},z)/\omega_{{\bf q}})$.
Moreover $\Sigma({\bf q},z)/\omega_{{\bf q}}$ is presented on a scale
of $\lambda^{2}/m$. The latter quantity encodes the strength of the
magnetoelastic coupling. For cuprates with simple spin super-exchange,
estimates of the latter exist\cite{Chernyshev2015}. For SOC assisted
Mott-insulators with pseudo-spin compass exchange of the Kitaev type,
this is an open issue and not part of our analysis. Fourth, we confine
the discussion to the imaginary part of $\Sigma({\bf q},\omega+i0^{+})$.
This is no loss of information, because of the Kramers-Kronig relation.
Fifth, performing the average over gauge configurations in Eq. (\ref{eq:12}),
we use an additional averaging, namely over periodic and antiperiodic
boundary conditions. This reduces finite size effects. Sixth and conceptually
important, our results for $\mathrm{Im}\Sigma({\bf q},\omega+i0^{+})$
are sixfold symmetric regarding the direction of ${\bf q}$. This
seems clear from the original spin hamiltonian and is obviously satisfied
in the homogeneous gauge. However also for $T\gtrsim T^{\star}$,
with random gauge links along the $z$-bonds, and for all momenta
${\bf q}$ considered, we find that $\mathrm{Im}\Sigma({\bf q},\omega+i0^{+})$
obeys this symmetry.

Now we consider the low-$T$ behavior, using the homogeneous gauge
ground state. A typical small-${\bf q}$ spectrum of $\Sigma({\bf q},\omega{+}i0^{+})$
is shown in Fig. \ref{fig:1}. From Eq. (\ref{eq:8}), it comprises
two decay types for the phonon, (i) a particle-hole (ph) and (ii)
a two-particle (pp) channel. On the scale of the plot only the latter
is visible. The inset refers to the ph-channel. In stark contrast
to usual phonon-electron scattering, the Fermi volume shrinks to zero
in the Kitaev model as $T{\rightarrow}0$, i.e., occupied states only
stem from a small patch with $\epsilon_{{\bf k}}\lesssim T$ around
the Dirac cone. Therefore, the weight of the ph-channel decreases
rapidly to zero as $T{\rightarrow}0$. In this regime and for small-${\bf q}$,
because of the linear fermion dispersion close to the cones, the spectral
support of the ph-continuum is roughly confined to a narrow strip
of order $\omega\in[\max(0,\epsilon_{{\bf q}}-2T),\epsilon_{{\bf q}}]$.
At the upper edge of this continuum the ph DOS is singular. The inset
of Fig. \ref{fig:1} is consistent with this, considering the finite
system size and imaginary broadening used. Regarding the pp-channel,
the complete two-particle continuum is unoccupied and available for
intermediate states as $T{\rightarrow}0$. This leads to the broad
spectral hump seen in Fig. \ref{fig:1}, which extends out to $\max(2\epsilon_{{\bf k}})=3J_{z}$,
at $J_{x,y}=J_{z}$ and is two orders of magnitude larger than the
ph-process at this temperature.

We note that for systems with small-${\bf q}$ phonon velocities,
comparable to those of the fermions at the Dirac cone, the on-shell
phonon damping $\mathrm{Im}\Sigma({\bf q},\omega_{{\bf q}}{+}i0^{+})$
stems from an energy range similar to that of the inset in Fig. \ref{fig:1}.
In view of the strong suppression of the ph-channel, the low-$T$
phonon damping, if due to scattering from mobile matter fermions,
would primarily result from two-particle decay.

\begin{figure}[b]
\centering{}\includegraphics[width=0.98\columnwidth]{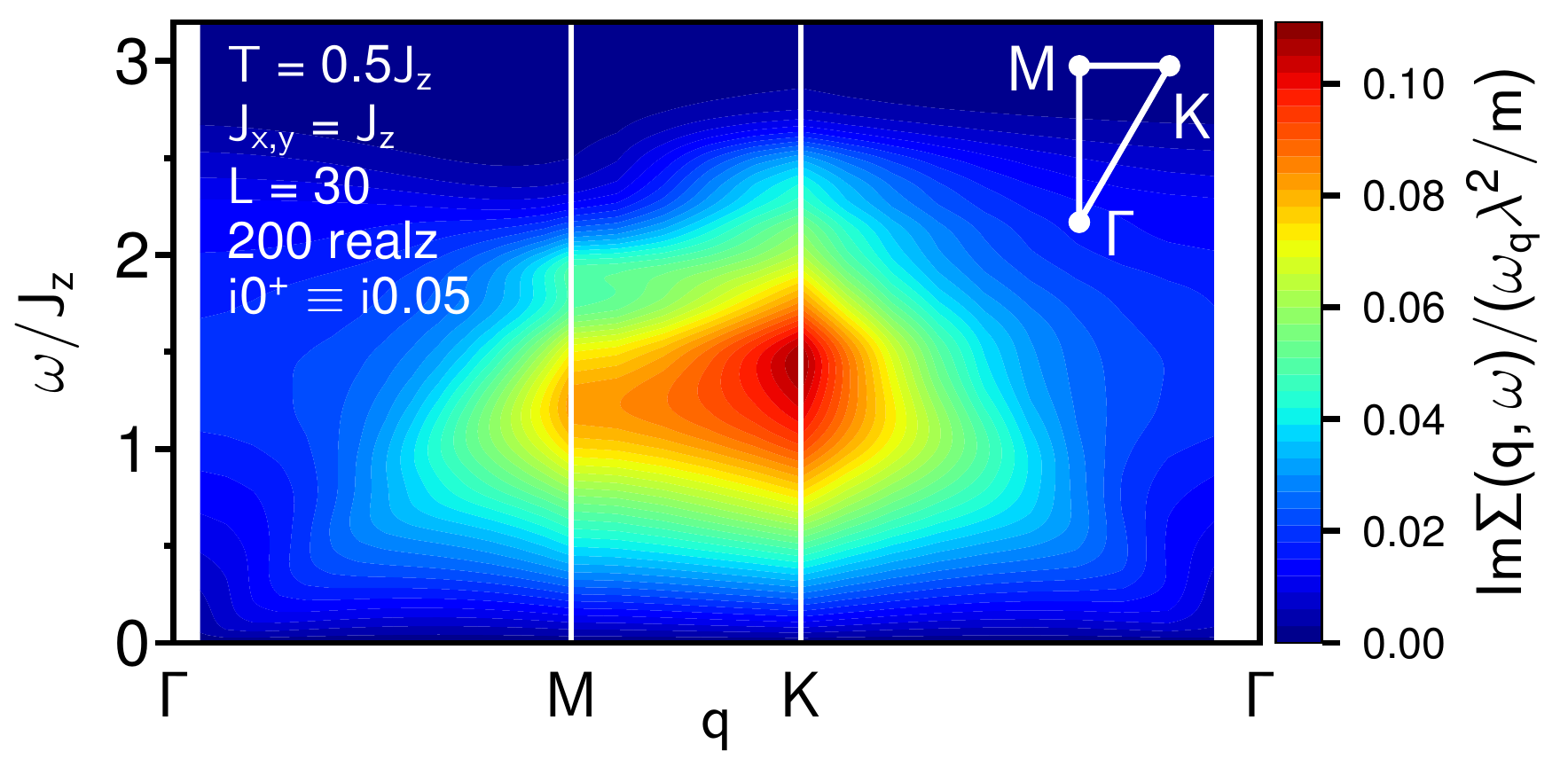}
\caption{Contours of the imaginary part $\mathrm{Im}\Sigma({\bf q},\omega+i0^{+})$
of the phonon self-energy at $T=0.5J_{z}\gtrsim T^{\star}$ in the
random gauge state versus ${\bf q},\omega$ along the path $\Gamma$-$M$-$K$-$\Gamma$
of the BZ.\label{fig:3}}
\end{figure}

Next we focus on temperatures above the flux proliferation, i.e. $T\gtrsim T^{\star}$,
using a random gauge state. Fig. \ref{fig:2} show the spectrum of
$\Sigma({\bf q},\omega{+}i0^{+})$ for three representative low-${\bf q}$
values versus $\omega$. Decomposing Eq. (\ref{eq:12}) into addends
with $\epsilon_{\sigma}\epsilon_{\rho}\gtrless0$ the ph- and pp-contributions
to $\mathrm{Im}\Sigma({\bf q},\omega{+}i0^{+})$ can be extracted
and are also shown. For comparison, the spectrum for completely identical
system parameters, however in the absence of gauge disorder, i.e. for a homogeneous ground state gauge
is included. Small oscillations in the latter are due to larger finite
size effects within the homogeneous gauge. Fig. \ref{fig:2} highlights
the drastic impact of thermally excited fluxes. While \emph{quantitatively},
keeping a homogeneous gauge, elevated temperatures merely increase
the weight of the ph-channel, \emph{qualitatively} the latter remains
a narrow structure below ${\sim}\epsilon_{{\bf q}}$ in the small-${\bf q}$
limit. This situation changes completely in the thermally excited
gauge background. As is obvious from the figure, the ph-channel spreads
into a broad feature, extending over roughly the entire one-particle
energy range. The shape of this feature is modulated by ${\bf q}$.
The pp-channel on the other hand seems less affected by the gauge
disorder, with a shape qualitatively similar to  that in the gauge
ground state.

Interestingly, these findings bear some resemblance to studies of
the dynamical thermal conductivity $\kappa(\omega)$ in Kitaev spin
systems\cite{Metavitsiadis2017}. While this is a completely different
${\bf q}=0$ correlation function, it also displays a sharp low frequency 
structure, the so-called Drude-peak $\sim\delta(\omega)$ and a pp-continuum in the homogeneous
gauge. However for $T\gtrsim T^{\star}$ the Drude peak is smeared
over an energy range $\sim J_{x,y,z}$ by fermions scattering from
thermally excited gauges, while the pp-continuum is less affected.
Here, Fig. \ref{fig:2} signals a small systematic reduction of the
fermion band-width for $T\gtrsim T^{\star}$. Again, the same effect
is found in $\kappa(\omega)$.

In Fig. \ref{fig:3} the spectrum of the phonon self-energy is displayed
along a path connecting high-symmetry points in the BZ for $T\gtrsim T^{\star}$.
This figure has to be taken with a grain of salt. As has been emphasized,
Eq. (\ref{eq:3}) is an approximation for the long wave-length limit,
neglecting the two-site basis of the honeycomb lattice regarding the
phonons. Therefore the large-${\bf q}$ spectra in Fig. \ref{fig:3}
are approximate only. Our main point however, should remain unaffected
by this. Namely, that while the two distinct types of relaxation channels,
i.e. ph and pp, contained in the low-$T$ limit Eq. (\ref{eq:8})
and the sharp ph-spike in the spectra assuming a homogeneous gauge,
might suggest that $\mathrm{Im}\Sigma({\bf q},\omega{+}i0^{+})$ should
display a dispersive feature, shadowing the fermion dispersion, this
is \emph{not} so. On the contrary, for $T\gtrsim T^{\star}$ and because
of thermally excited gauges $\mathrm{Im}\Sigma({\bf q},\omega{+}i0^{+})$
in Fig. \ref{fig:3} is almost featureless, covering all of the energy
range $\omega\in[0,\max(2\epsilon_{{\bf k}})]$ at any momentum. At
low-${\bf q}$, remnants of the fermion ph-continuum boundary can
be observed dispersing upwards, however this is only a weak phenomenon.
For ${\bf q}$ approaching the $M$ and $K$ points there is a global
intensity increase. The role of the small-${\bf q}$ approximation
for this is unclear. Finally, for $T\lesssim T^{\star}$ the spectrum
is also weakly dispersive only, because of the strong suppression
of the ph-channel, as discussed in Fig. \ref{fig:1}.

\begin{figure}[tb]
\centering{}\includegraphics[width=0.8\columnwidth]{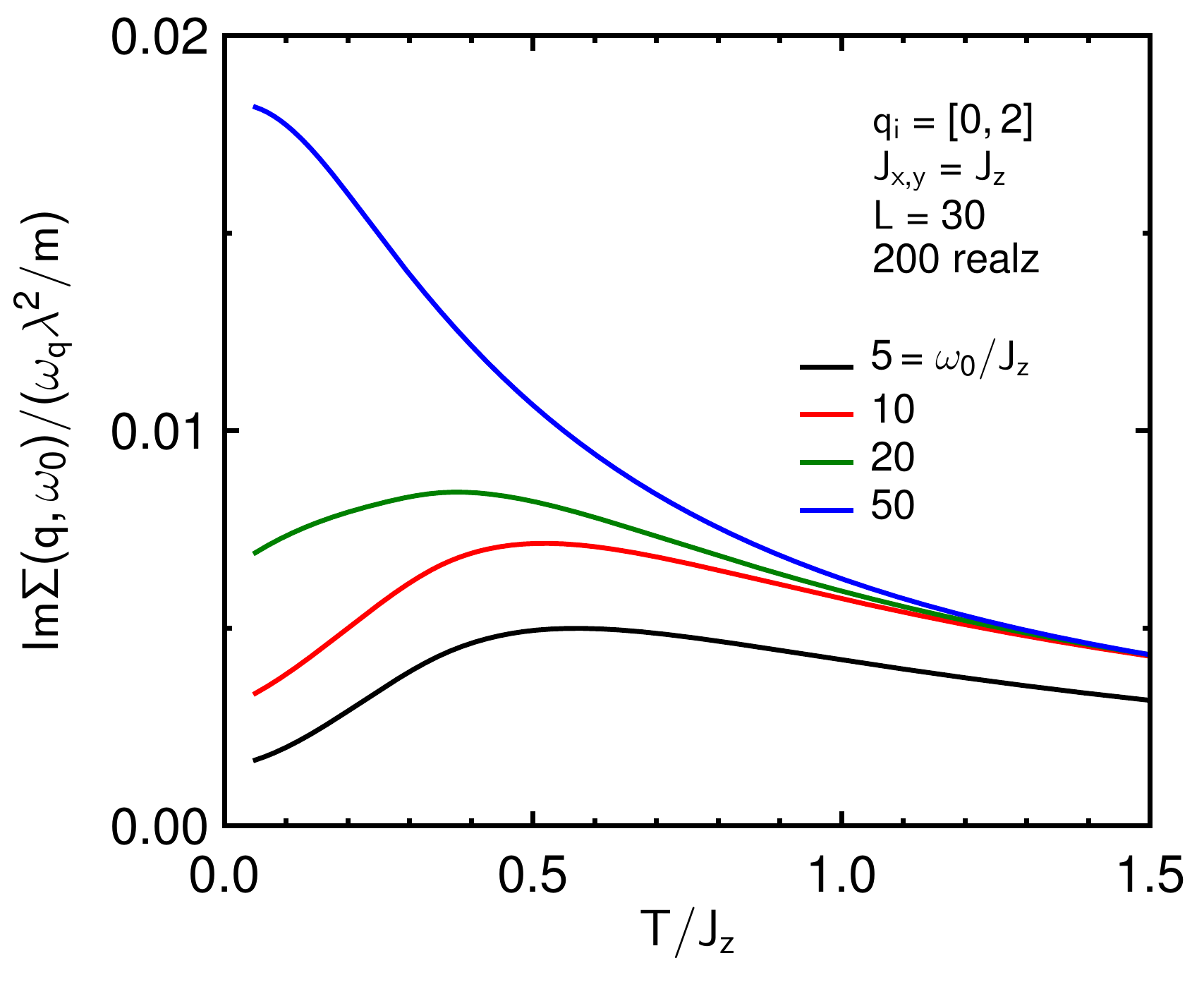}
\caption{$\mathrm{Im}\Sigma({\bf q},\omega_{0}+i0^{+})$ versus $T$ for fixed
momentum ${\bf q}=2\pi/L\sum_{j=1,2}q_{ij}{\bf G}_{j}$ and four energies
$\omega_{0}$, using a random gauge state and $T\geqslant0.05J_{z}$.\label{fig:4}}
\end{figure}

Finally we turn to the temperature dependence of the phonon lifetime.
This requires realistic dispersions $\omega_{{\bf q}}$ and values
for $\lambda^{2}/m$ to either solve for $z^{2}-\omega_{{\bf q}}^{2}-2\omega_{{\bf q}}\Sigma({\bf q},z)=0$
self consistently, or approximately use the on-shell self-energy $\mathrm{Im}\Sigma({\bf q},\omega_{{\bf q}}{+}i0^{+})$.
Lacking this information, we nevertheless consider the latter quantity
for fixed values of ${\bf q},\omega$ using three potentially 'typical'
acoustic phonon energies $\omega$ for the chosen low-${\bf q}$ wave
vector. This is shown in Fig. \ref{fig:4}. 
 The figure highlights two points. First, the qualitative variations
of the phonon lifetime with $T$ strongly depends on the actual phonon
energy. Second, and universally, since the phonons scatter off a reservoir
of fermions, they will \emph{undress} as the latter turns classical,
i.e. as $T\gg J_{x,y,z}$. Since $J_{x,y,z}$ in several Kitaev materials
can be of order of the Debye energy, this may be of experimental relevance. 

\section{Summary\label{sec:Summary}}

In conclusion, phonons in Kitaev magnets experience scattering from
a characteristic continuum of excitations comprising fractionalized
fermionic quasiparticles, allowing for ph- as well as pp-decay channels.
In contrast to conventional phonon-electron coupling, thermally excited
random gauge fluxes strongly broaden the low-energy ph-channel and
render the scattering only weakly dispersive. Our study remains with
several open questions. These include a more realistic treatment of
the phonon polarizations and band structure, a microscopic analysis
of the strength of the magnetoelatic coupling, and related to that,
the role of vertex corrections. Moreover, other types of phonon-(pseudo)spin
interactions should be considered. Finally, regarding the impact of non-Kitaev
exchange, it is tempting to speculate, that similar to previous analysis
of transport\cite{Metavitsiadis2019a}, the pp-decay channel will persist
to signal remnants of Majorana physics in the phonon renormalization, even
if a sizable Heisenberg exchange is added to the model.

\medskip{}

\emph{Acknowledgments}: We thank R. Valenti and C. Hess for discussion,
moreover we are grateful to R. Valenti and S. Biswas for providing
us with visualization data\cite{privateValenti2019} of lattice vibrations
in $\alpha$-RuCl$_{3}$. Work of W.B. has been supported in part
by the DFG through project A02 of SFB 1143 (project-id 247310070),
by Nds. QUANOMET, and the NTH-School CiN. W.B. also acknowledges kind
hospitality of the PSM, Dresden. This research was supported in part
by the National Science Foundation under Grant No. NSF PHY-1748958.


\begin{thebibliography}{10}
\bibitem{Balents2010}L. Balents, Nature \textbf{464}, 199-208 (2010).

\bibitem{Balents2016}L. Savary and L. Balents, Rep. Prog. Phys. \textbf{80},
016502 (2017). 

\bibitem{Kitaev2006}A. Kitaev, Ann. Phys. (N.Y.) \textbf{321}, 2
(2006).

\bibitem{Feng2007}X.-Y. Feng, G.-M. Zhang, and T. Xiang, Phys. Rev.
Lett. \textbf{98}, 087204 (2007).

\bibitem{Chen2008}H.-D. Chen and Z. Nussinov, J. Phys. A: Math. Theor.
\textbf{41}, 075001 (2008).

\bibitem{Nussinov2009}Z. Nussinov and G. Ortiz, Phys. Rev. B \textbf{79},
214440 (2009).

\bibitem{Mandal2012}S. Mandal, R. Shankar and G. Baskaran, J. Phys.
A: Math. Theor. \textbf{45}, 335304 (2012).

\bibitem{Khaliullin2005}G. Khaliullin, Prog. Theor. Phys. Suppl.
\textbf{160}, 155 (2005).

\bibitem{Jackeli2009}G. Jackeli and G. Khaliullin, Phys. Rev. Lett.
\textbf{102}, 017205 (2009). 

\bibitem{Chaloupka2010}J. Chaloupka, G. Jackeli, and G. Khaliullin
Phys. Rev. Lett. \textbf{105} 027204 (2010).

\bibitem{Nussinov2015}Z. Nussinov and J. van den Brink, Rev. Mod.
Phys. \textbf{87}, 1 (2015).

\bibitem{Trebst2017}S. Trebst, \emph{Kitaev Materials}, Lecture Notes
of the 48th IFF Spring School 2017, S. Blügel, Y. Mokrousov, T. Schäpers,
Y. Ando (Eds.), ISBN 978-3-95806-202-3

\bibitem{Banerjee2016}A. Banerjee, C. A. Bridges, J. Q. Yan, A. A.
Aczel, L. Li, M. B. Stone, G. E. Granroth, M. D. Lumsden, Y. Yiu,
J. Knolle, S. Bhattacharjee, D. L. Kovrizhin, R. Moessner, D. A. Tennant,
D. G. Mandrus, and S. E. Nagler, Nat. Mater. \textbf{15}, 733 (2016).

\bibitem{Banerjee2016a}A. Banerjee, J. Yan, J. Knolle, C. A. Bridges,
M. B. Stone, M. D. Lumsden, D. G. Mandrus, D. A. Tennant, R. Moessner,
and S. E. Nagler, Science \textbf{356}, 6342 (2017).

\bibitem{Banerjee2018}A. Banerjee, P. Lampen-Kelley, J. Knolle, C.
Balz, A. A. Aczel, B. Winn, Y. Liu, D. Pajerowski, J. Yan, C. A. Bridges,
A. T. Savici, B. C. Chakoumakos, M. D. Lumsden, D. A. Tennant, R.
Moessner, D. G. Mandrus, and S. E. Nagler, Nat. Part. J. Quantum Mater.
\textbf{3}, 8 (2018).

\bibitem{Knolle2014}J. Knolle, G.W. Chern, D.L. Kovrizhin, R. Moessner,
and N.B. Perkins, Phys. Rev. Lett. \textbf{113}, 187201 (2014).

\bibitem{Baek2017}S.-H. Baek, S.-H. Do, K. Y. Choi, Y.S. Kwon, A.U.B.
Wolter, S. Nishimoto, J. van den Brink, and B. Büchner, Phys. Rev.
Lett. \textbf{119}, 037201 (2017). 

\bibitem{Zheng2017}J. Zheng, K. Ran, T. Li, J. Wang, P. Wang, B.
Liu, Z.-X. Liu, B. Normand, J. Wen, and W. Yu, Phys. Rev. Lett. \textbf{119},
227208 (2017). 

\bibitem{Plumb2014}K. W. Plumb, J. P. Clancy, L. J. Sandilands, V.
V. Shankar, Y. F. Hu, K. S. Burch, H.-Y. Kee, and Y.-J. Kim, Phys.
Rev. B \textbf{90}, 041112 (2014).

\bibitem{Kasahara2018}Y. Kasahara, T. Ohnishi, Y. Mizukami, O. Tanaka,
S. Ma, K. Sugii, N. Kurita, H. Tanaka, J. Nasu, Y. Motome, T. Shibauchi,
and Y. Matsuda, Nature \textbf{559}, 227 (2018).

\bibitem{Cookmeyer2018}J. Cookmeyer, and J. E. Moore, Phys. Rev.
B \textbf{98}, 060412(R) (2018).

\bibitem{McClarty2018}P. A. McClarty, X. -Y. Dong, M. Gohlke, J.
G. Rau, F. Pollmann, R. Moessner, and K. Penc, Phys. Rev. B \textbf{98},
060404(R) (2018).

\bibitem{Ye2018}M. Ye, G. B. Halász, L. Savary, and L. Balents, Phys.
Rev. Lett. \textbf{121}, 147201 (2018).

\bibitem{Vinkler-Aviv2018}Y. Vinkler-Aviv, and A. Rosch, Phys. Rev.
X \textbf{8}, 031032 (2018). 

\bibitem{Hirobe2017}D. Hirobe, M. Sato, Y. Shiomi, H. Tanaka, and
E. Saitoh, Phys. Rev.\textasciitilde{}B 95, 241112 (2017).

\bibitem{Leahy2017}I.A. Leahy, C.A. Pocs, P.E. Siegfried, D. Graf,
S.-H. Do, K.-Y. Choi, B. Normand, and M. Lee, Phys. Rev. Lett. \textbf{118},
187203 (2017).

\bibitem{Hentrich2018}R. Hentrich, A. U. B. Wolter, X. Zotos, W.
Brenig, D. Nowak, A. Isaeva, T. Doert, A. Banerjee, P. Lampen-Kelley,
D. G. Mandrus, S. E. Nagler, J. Sears, Y.-J. Kim, B. Büchner, C. Hess,
Phys. Rev. Lett. \textbf{120}, 117204 (2018).

\bibitem{Yu2018}Y. J. Yu, Y. Xu, K. J. Ran, J. M. Ni, Y. Y. Huang,
J. H. Wang, J. S. Wen, and A. Y. Li, Phys. Rev. Lett. \textbf{120},
067202 (2018).

\bibitem{Glamazda2017}A. Glamazda, P. Lemmens, S.-H. Do, Y. S. Kwon,
and K.-Y. Choi, Phys. Rev. B \textbf{95}, 174429 (2017). 

\bibitem{Sahasrabudhe2019}A. Sahasrabudhe, D. A. S. Kaib, S. Reschke,
R. German, T. C. Koethe, J. Buhot, D. Kamenskyi, C. Hickey, P. Becker,
V. Tsurkan, A. Loidl, S. H. Do, K. Y. Choi, M. Grüninger, S. M. Winter,
Z. Wang, R. Valenti, and P. H. M. van Loosdrecht, \emph{ArXiv:1908.11617
{[}Cond-Mat{]}} (2019). 

\bibitem{Reschke2019}S. Reschke, V. Tsurkan, S.-H. Do, K.-Y. Choi,
P. Lunkenheimer, Z. Wang, and A. Loidl, Phys. Rev. B \textbf{100},
100403(R) (2019). 

\bibitem{Widmann2019}S. Widmann, V. Tsurkan, D. A. Prishchenko, V.
G. Mazurenko, A. A. Tsirlin, and A. Loidl, Phys. Rev. B \textbf{99},
094415 (2019).

\bibitem{Biesner2018}T. Biesner, S. Biswas, W. Li, Y. Saito, A. Pustogow,
M. Altmeyer, A. U. B. Wolter, B. Büchner, M. Roslova, T. Doert, S.
M. Winter, R. Valentí, and M. Dressel, Phys. Rev. B 97, 220401 (2018).

\bibitem{Bastien2018}G. Bastien, G. Garbarino, R. Yadav, F. J. Martinez-Casado,
R. Beltrán Rodríguez, Q. Stahl, M. Kusch, S. P. Limandri, R. Ray,
P. Lampen-Kelley, D. G. Mandrus, S. E. Nagler, M. Roslova, A. Isaeva,
T. Doert, L. Hozoi, A. U. B. Wolter, B. Büchner, J. Geck, and J. van
den Brink, Phys. Rev. B 97, 241108 (2018). 

\bibitem{Yadav2018}R. Yadav, S. Rachel, L. Hozoi, J. van den Brink,
and G. Jackeli, Phys. Rev. B 98, 121107 (2018).

\bibitem{Stamokostas2017}G. L. Stamokostas, P. E. Lapas, and G. A.
Fiete, Phys. Rev. B \textbf{95}, 064410 (2017).

\bibitem{Nasu2015}J. Nasu, M. Udagawa, and Y. Motome, Phys. Rev.
B \textbf{92}, 115122 (2015). 

\bibitem{Metavitsiadis2017}A. Metavitsiadis, A. Pidatella, and W.
Brenig, Phys. Rev. B \textbf{96}, 205121 (2017).

\bibitem{Pidatella2019}A. Pidatella, A. Metavitsiadis, and W. Brenig
Phys. Rev. B \textbf{99}, 075141 (2019).

\bibitem{Metavitsiadis2017a}A. Metavitsiadis and W. Brenig, Rev.
B \textbf{96}, 041115(R) (2017).

\bibitem{AGD}A. A. Abrikosov, L. P. Gorkov, and I. E. Dzyaloshinski,
\emph{Methods of Quantum Field Theory in Statistical Physics}, Dover
Publications Inc., ISBN 0486140156

\bibitem{Migdal1958} A. B. Migdal, Sov. Phys. JETP \textbf{7},
996 (1958).

\bibitem{Roy2014}B. Roy, J. D. Sau, and S. Das Sarma, Phys. Rev.
B 89, 165119 (2014).

\bibitem{Chernyshev2015}A. L. Chernyshev and W. Brenig, Phys. Rev.
B \textbf{92}, 054409 (2015).

\bibitem{privateValenti2019}R. Valenti, S Biswas, private communication,
\emph{unpublished}

\bibitem{Metavitsiadis2019a}
A. Metavitsiadis, C. Psaroudaki, and W. Brenig, Phys. Rev. B
{\bf 99}, 205129 (2019).


\end{thebibliography}
\end{document}